\documentclass[namedreferences]{SolarPhysics}
\usepackage[optionalrh]{spr-sola-addons} 
\usepackage{graphicx}        
\usepackage{color}           
\usepackage{url}             

\newcommand{\aap}{{\it Astron. Astrophys.} }

\newcommand{\apj}{{\it Astrophys. J.} }
\newcommand{\apjl}{{\it Astrophys. J.} }

\newcommand{\nat}{{\it Nature} }

\newcommand{\solphys}{{\it Solar Phys.} }

\newcommand{\bk}{ \mbox{\boldmath$k$} }

\newcommand{\bvel}{{\mbox{\boldmath$v$}}}

\newcommand{\ms}{m\,s$^{-1}$}

\begin{document}

\begin{article}

\begin{opening}

\title{Observation and Modeling of the Solar-Cycle Variation of the Meridional Flow}

\author{Laurent~\surname{Gizon}$^{1}$\sep
        Matthias~\surname{Rempel}$^{2}$
       }
\runningauthor{L.~Gizon, M.~Rempel}
\runningtitle{Solar-Cycle Variation of the Meridional Flow}

\institute{$^{1}$ Max-Planck-Institut f\"{u}r Sonnensystemforschung, 37191 Katlenburg-Lindau, Germany \email{gizon@mps.mpg.de}\\
              $^{2}$ High Altitude Observatory, National Center for Atmospheric Research, P.O. Box 3000, Boulder, CO 80307, USA}

\date{Received: 7 September 2007  /  Accepted: }

\begin{abstract}
We present independent observations of the solar-cycle variation of  flows near the solar surface and at a depth of about 60~Mm, in the latitude range $\pm45^\circ$. We show that the time-varying components of the meridional flow at these two depths have opposite sign, while the time-varying components of the zonal flow are in phase. This is in agreement with previous results. We then investigate whether the observations are consistent with a theoretical model of solar-cycle dependent meridional circulation  based on a flux-transport dynamo combined with a geostrophic flow caused by increased radiative loss in the active region belt (the only existing quantitative model). We find that the model and the data are in qualitative agreement, although the amplitude of the solar-cycle variation of the meridional flow at 60~Mm is underestimated by the model.  
\end{abstract}

\end{opening}

\section{Introduction}
Solar oscillations are a unique tool to infer conditions inside the Sun. They have been recorded with extreme precision since 1996 with the Michelson Doppler Imager (MDI, \opencite{Scherrer1995}) onboard the Solar and Heliospheric (SOHO) space telescope. Large-scale rotation inside the Sun can be estimated by inversion of the frequencies of millions of global modes of oscillation ({\it e.g.}, \opencite{Schou1998}). Rotation is known to vary with time in the solar interior at the level of about $\pm10$~\ms ({\it e.g.}, \opencite{Schou1999}; \opencite{Howe2000}; \opencite{Vorontsov2002}; \opencite{Howe2006a}; \opencite{Howe2006b}). These variations, known as torsional oscillations, consist of bands of faster and slower rotation that migrate in latitude as the eleven-year solar magnetic cycle develops.  Torsional oscillations may be driven by the Lorentz force due to a dynamo wave \cite{Schussler1981, Yoshimura1981, Covas2000}. Other explanations have been proposed (see \opencite{Shibahashi2004} and references therein),  
including the suggestion by \inlinecite{Spruit2003} that torsional oscillations are driven by horizontal pressure gradients caused by photospheric magnetic activity. 

Long time averages of surface Doppler measurements have shown the existence of a flow from the equator to the poles with an amplitude of 10\,--\,20~\ms. 
An introduction to the theory of solar meridional circulation is provided
by \inlinecite{Shibahashi2007}. 
Temporal variations in the meridional flow have been reported by several authors.  By tracking the small photospheric magnetic features, \inlinecite{Komm1993b}, \inlinecite{Komm1994}, and \inlinecite{Meunier1999} found a significant change in the meridional flow near sunspot latitudes, implying a solar-cycle variation.  Variations in the surface Doppler meridional velocity have been detected by \inlinecite{Ulrich2005}, in particular at latitudes above $60^\circ$.

Local helioseismology (see, {\it e.g.}, \opencite{Gizon2005}) also provides reasonable measurements of the meridional circulation for latitudes below about $50^\circ$ \cite{Giles1997}. The time-varying component of the meridional flow with respect to a long-term average does not exceed $\pm5$~\ms and is consistent with a small near-surface inflow toward active latitudes \cite{Basu2003, Gizon2004, Zhao2004, Gonzalez2006, Komm2006} and an outflow from active latitudes at depths greater than 20~Mm \cite{Chou2001,Beck2002,Chou2005}. As discussed by \inlinecite{Gizon2004}, these variations would appear to be caused by localized flows around localized regions of magnetic activity \cite{Gizon2001,Haber2001, Haber2004}. 

The purpose of this paper is two-fold. First, we provide independent measurements of the temporal variations of the meridional circulation near the solar surface and at a depth of about 60~Mm.  For the near-surface meridional flow, we use an original technique, which consists of measuring the advection of the supergranulation pattern \cite{Gizon2003}. The deeper meridional flow is calibrated from an earlier time-distance helioseismology observation by \inlinecite{Beck2002}. The meridional circulation measurements at these two depths are compared, in particular by looking at the eleven-year periodicity of the flows (Section \ref{sec.obs}). Our study of meridional circulation confirms previous observations (listed above) and is complementary to the analysis of zonal flows by \inlinecite{Howe2006a}. 

Second, we wish to compare the observations with a theoretical model \cite{Rempel2005a,Rempel2005b,Rempel2006} based on a flux-transport dynamo combined with a geostrophic flow caused by increased radiative loss in the active region belt, according to Spruit's (2003) original idea. Section~\ref{sec.model} provides a description of the model with a focus on meridional flows, since this aspect of the model has not been discussed elsewhere in detail. To our knowledge, this model is the only existing quantitative model of the solar-cycle dependence of internal flows: it is natural to ask whether it is consistent with the observations.   The comparison between the observations and the model (Section \ref{sec.discussion}) is encouraging, although some inconsistencies cannot be ignored.

\section{Observations of the Meridional Flow}
\label{sec.obs}

\subsection{Near-Surface Layers}
\label{sec.obssurf}
The method we employ to infer flows near the solar surface is based on the analysis of \inlinecite{Gizon2003}, which was originally applied to a single data set from 1996. Here we use a series of MDI full-disk Doppler images covering the period 1996\,--\,2002. Each year, two to three months of continuous Dopplergrams are available for analysis (MDI Dynamics runs). The MDI data after 2002 were not used simply because we are analyzing an existing pre-processed data set.

Dopplergrams were tracked at the Carrington rotational velocity to remove the main component of rotation. We used {\it f}-mode time-distance helioseismology \cite{Duvall2000} to obtain every 12~hour a $120^\circ\times120^\circ$ map of the horizontal divergence of the flow field $1$~Mm below the photosphere. The main component of the divergence signal is due to supergranulation. For any given target latitude ($\lambda$) we considered a longitudinal  section of the data $10^\circ$ wide in latitude. Using a local plane-parallel approximation in the neighborhood of latitude $\lambda$, the divergence signal was interpolated onto a Cartesian grid sampled at $2.92$~Mm in the $x$ (prograde) and $y$ (northward) coordinates. The divergence signal was decomposed into its harmonic components $\exp({\rm i}(k_x x + k_y y - \omega t))$ to obtain a local power spectrum $P(\bk,\omega;\lambda)$, where $\bk=(k_x,k_y)$ is the horizontal wavevector and $\omega$ is the angular frequency. At fixed $kR_\odot=120$, we fit for two functions $f$ and $g$ and a horizontal vector $\bvel$ such that $P(\bk,\omega; \lambda) = f(\bk) \, g(\omega - \bk\cdot\bvel(\lambda))$. This representation fits the data adequately.  As was done by \inlinecite{Gizon2003}, we interpret $\bvel(\lambda)$ to be a horizontal flow causing a Doppler shift $\Delta\omega=\bk\cdot\bvel$. This flow is likely to be an average over the supergranulation layer, which has been estimated to reach depths greater than 10~Mm by \inlinecite{Zhao2003}. 

\begin{figure}[]
\centering
\includegraphics[width=11.cm]{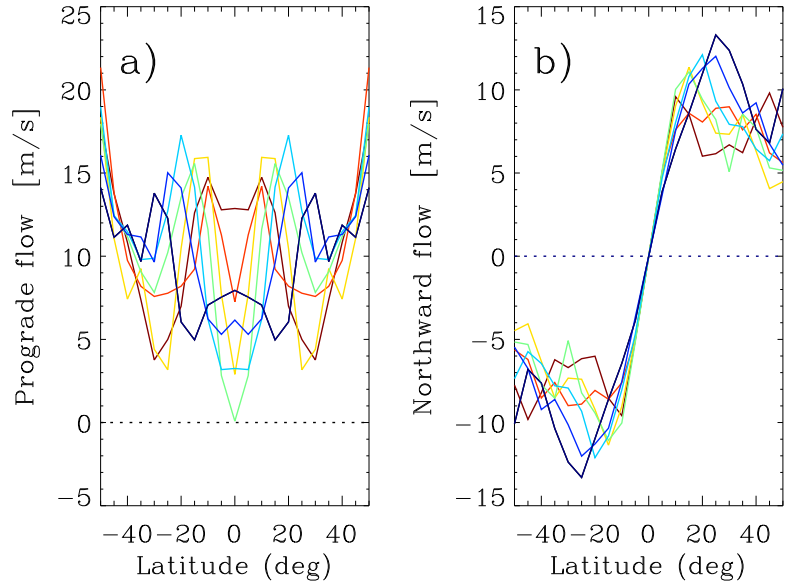}
\caption{
(a) Rotational velocity ($v_x$) and (b) meridional flow ($v_y$) near the solar surface as a function of latitude ($\lambda$). Each MDI dynamics run is plotted with a different color from blue in 1996 to red in 2002. The rotational velocity is given with respect to the rotational velocity of the small magnetic features (Komm, Howard, and Harvey, 1993a). }
\label{fig.surface}
\end{figure}

Figure~\ref{fig.surface} is a plot of $v_x(\lambda)$ and $v_y(\lambda)$ for each full-disk MDI run as a function of latitude in the range $|\lambda|<50^\circ$. To reduce random noise, the North\,--\,South symmetric component of $v_x$ and the antisymmetric component of $v_y$ are extracted. Over the period 1996\,--\,2002, $v_x$ varies by $12$~\ms peak-to-peak at the Equator (Figure \ref{fig.surface}a). The Meridional flow is poleward with a mean amplitude of 10~\ms at latitude $20^\circ$ (Figure \ref{fig.surface}b). The peak-to-peak variation of the meridional flow is 7~\ms at $\lambda=30^\circ$, {\it i.e.}~a significant fraction of the time-average value. We estimate that the standard deviation of the noise at a particular latitude ($5^\circ$ bin) for any given year is less than 1~\ms. The systematic errors that depend on position on the solar disk have been measured to be very low (less than $5$~\ms over the $120^\circ\times120^\circ$ region of analysis).

\subsection{Deeper Inside the Sun}
\label{sec.obsdeep}
In order to probe deeper layers into the solar convection zone, we used acoustic waves and time-distance helioseismology. For each three-month period, travel times were measured by cross-correlation of the Doppler oscillation signal recorded during the MDI structure program (nearly continuous coverage but lower spatial resolution) according to the procedure described by \inlinecite{Giles1999}. Using a mean travel distance of $17^\circ$ enables us to probe layers about $60$~Mm below the surface. The full details of this analysis can be found in \inlinecite{Beck2002}. Waves that propagate in the North\,--\,South direction are used to infer the meridional flow, while waves that propagate East\,--\,West are used to infer the zonal flows. In order to convert travel-time shifts into flows in units of \ms, we use a simple calibration of $v_x$ at a depth based on the observation by \inlinecite{Howe2006a} (global-mode helioseismology) that the amplitude of the time-varying component of the zonal flow is nearly independent of depth. We choose the near-surface zonal-flow measurements of Section \ref{sec.obssurf} as a reference. The calibration of $v_x$ is then used to calibrate $v_y$. We find that the meridional flow at a depth of 60~Mm is poleward at all latitudes and has a maximum value of 6~\ms at latitude $25^\circ$. For a particular year and at fixed latitude ($5^\circ$ bin),  the standard deviation of the noise is about 2~\ms, significantly more than for the near-surface measurements.  

\subsection{Solar-Cycle Variations}
In order to discuss the solar-cycle dependence of the flows and to study the phase relationship between the flows measured at the two different depths, we extract the eleven-year periodic component from the data, as was done by \inlinecite{Vorontsov2002} and \inlinecite{Howe2006a} for zonal flows. At each latitude $\lambda$ and for each depth, we fit a function of the form
\begin{equation}
\tilde{v}_i (\lambda,t) = \overline{v}_i (\lambda) + v_i'(\lambda)  \, \cos\left( \frac{2\pi t}{11\,{\rm yr}}  + \phi_i(\lambda) \right) 
\end{equation}
to the observed velocity $v_i(\lambda,t)$, where the index $i$ refers to either the $x$ or the $y$ component of the flow. The long-term average is given by $\overline{v}_i$, while the amplitude and the phase of the time-varying component are denoted by $v_i'$ and $\phi_i$ respectively. 
We extract a 11-year periodicity from the data, since it is known from other observations that this is the dominant mode (we do not determine the 11-year periodicity based on the dataset itself). Shorter and longer periodicities are certainly present in the data; however, the length of the dataset does not allow for a determination of the full spectrum of modes.

\begin{figure}[]
\centering
\includegraphics[width=12.cm]{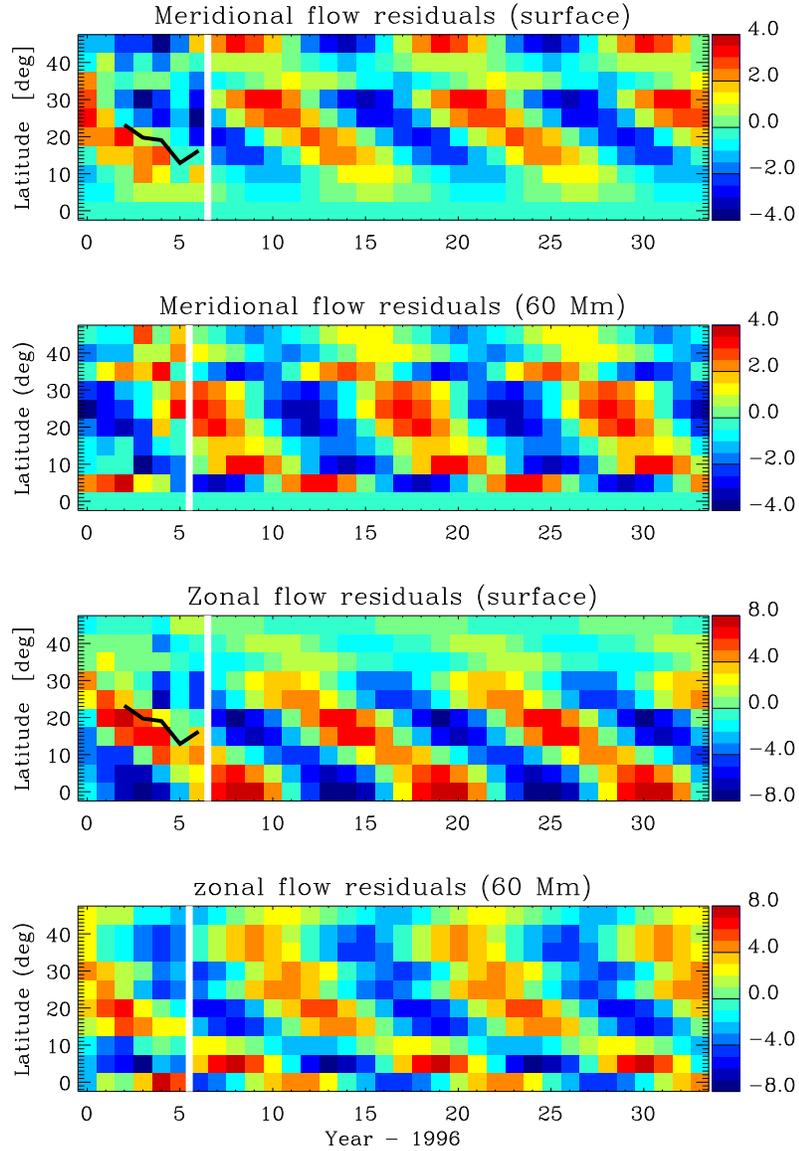}
\caption{Eleven-year periodic component of the meridional ({\it top two panels}) and zonal ({\it bottom two panels}) flows as function of time and latitude at two different depths in the solar interior: near the surface ({\it top and third panels}) and 60~Mm deep ({\it second and bottom panel}). The color bar is in units of \ms. 
A positive value indicates a poleward meridional or prograde zonal flow.
The observations ($v_i - \overline{v}_i$) cover the first six years, while the purely sinusoidal component ($\tilde{v}_i - \overline{v}_i$) is extrapolated in time (beyond the {\it white vertical white line}). The {\it black curves} indicate the mean latitude of magnetic activity.}
\label{fig.surf}
\end{figure}

\begin{figure}[]
\centerline{ \includegraphics[width=13cm]{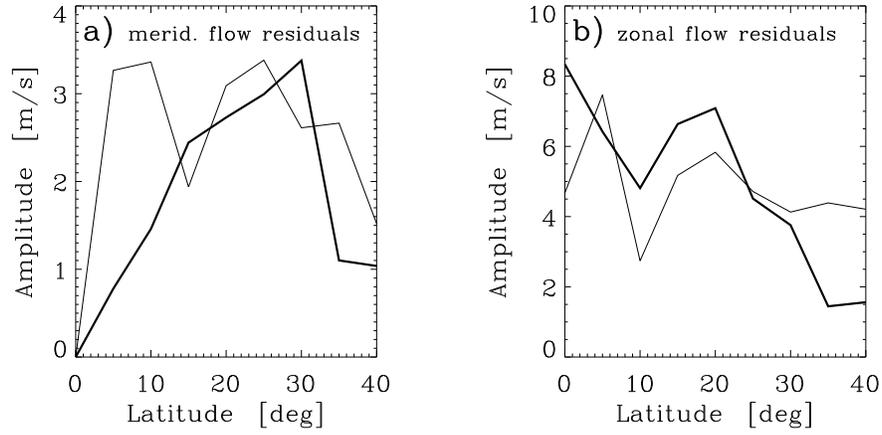} }
\caption{Amplitude (${v}_i'$) of the eleven-year periodic component of the (a)  meridional and (b) zonal flows. The near-surface values ({\it thick solid lines}) are absolute measurements. The calibration of the observations at 60~Mm depth ({\it thin lines}) follows from the assumption that the amplitude of the zonal torsional oscillation ({\it panel b}) is independent of depth over the latitude range $|\lambda|<45^\circ$.}
\label{fig.compare_ampli}
\end{figure}

\begin{figure}[]
\centering
\includegraphics[width=8cm]{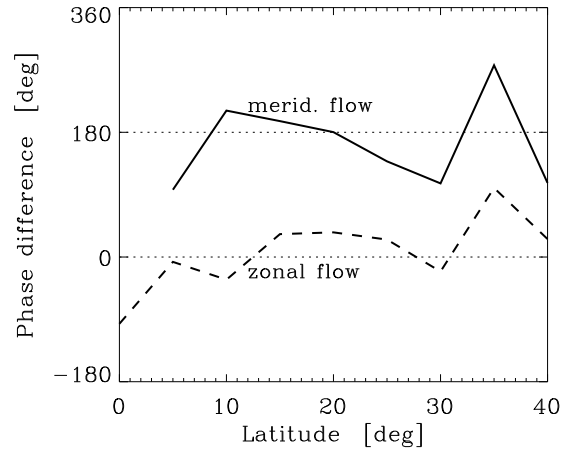}
\caption{Phase difference [$\Delta \phi = \phi({\rm deep}) - \phi({\rm surface})$] between the eleven-year periodic components of the flows measured at a depth of 60~Mm and near the surface. The {\it solid line} is for the meridional flow and the {\it dashed line} is for the zonal flow. 
}
\vspace{1.cm}
\label{fig.compare_phase}
\end{figure}

The eleven-year periodic components of the meridional and zonal flows are shown in Figure~\ref{fig.surf}. The torsional-oscillation pattern is clearly seen at both depths with an amplitude and a phase comparable to previous measurements ({\it e.g.}, \opencite{Howe2006a}).  The meridional flow also contains a significant eleven-year periodic component. Near the solar surface, the residuals indicate the presence of a North\,--\,South inflow toward the mean latitude of activity ({\it e.g.}, \opencite{Zhao2004}; \opencite{Komm2006}), while the data are consistent with a horizontal outflow from the mean latitude of activity deeper into the convection zone ({\it e.g.}, \opencite{Chou2001}; \opencite{Beck2002}).

Figure~\ref{fig.compare_ampli} gives the amplitudes of the time-varying components of the flows ($v_i'$). Under the assumption (Section \ref{sec.obsdeep}) that ${v}_x'$ does not vary appreciably with depth ($5$~\ms latitudinal average), then the amplitude of the time-varying meridional flow (${v}_y'$) is also found to be approximately independent of depth (${v}_y' \simeq 3$~\ms at $20^\circ$ latitude).   The evidence that the time-varying components of the meridional flow near the surface and deeper in the interior are anti-correlated is given in Figure~\ref{fig.compare_phase}, which shows the difference in $\phi_y$ at the two depths. On the contrary, there is no significant phase variation with depth for the zonal flow.

\section{Theoretical Model of Time-Varying Flows}
\label{sec.model}

The model results presented here are based on a non-kinematic flux-transport
dynamo model developed recently by Rempel. This model combines the differential rotation and meridional flow model of \inlinecite{Rempel2005a} with a flux-transport dynamo similar to the models of \inlinecite{Dikpati1999} and \inlinecite{Dikpati2001}. 
 We emphasize that this model is intended to give a fundamental 
understanding of the basic cycle properties and their relation to observable
variations of zonal and meridional flows. Therefore we focus here only on 
axisymmetric and North\,--\,South averaged quantities. Details of the model can be 
found in \inlinecite{Rempel2006}. Since a detailed comparison with observed 
torsional oscillations can be found in \inlinecite{Rempel2006} and \inlinecite{Howe2006b}, we focus here on the meridional flow variations.

The differential rotation model utilizes a meanfield Reynolds-stress approach that parametrizes the turbulent angular momentum transport (\opencite{Kitchatinov1993}; $\Lambda$-effect) leading to the observed equatorial acceleration. In this model the tachocline is forced through a uniform rotation boundary condition at the lower boundary of the computational domain. A meridional circulation, as required for a flux-transport dynamo, follows self-consistently through the Coriolis force resulting from the differential rotation. 

The computed differential rotation and meridional flow are used to advance the magnetic field in the flux-transport dynamo model, while the magnetic field is allowed to feed back through the meanfield Lorentz-force $\langle \mbox{\bf J}\rangle\times\langle\mbox{\bf B}\rangle$ (the contribution of the fluctuating part $\langle\mbox{\bf J}^{\prime}\times\mbox{\bf B}^{\prime}\rangle$ is not well known and neglected here).

We find in our model that the Lorentz-force feedback can only account for the
poleward propagating branch of the torsional oscillations, while the 
equatorward propagating branch in latitudes beneath $30^\circ$ requires 
additional physics. Parametrizing the idea proposed by \inlinecite{Spruit2003}
that the low-latitude torsional oscillation is a geostrophic flow caused
by increased radiative loss in the active region belt (due to small scale
magnetic flux) leads in our model to a surface oscillations pattern in good
agreement with observations. In order to force a torsional oscillation with
around $1$ nHz amplitude, a temperature variation of around $0.2$~K is
required. As a side effect the cooling produces close to the surface (in our 
model at $r=0.985\,R_{\odot}$) an inflow into the active-region belt of around
$2.3$~\ms.

We incorporated this process by adding a 
surface-cooling term that is dependent on the toroidal-field strength at the 
base of the convection zone, which is assumed to be the source for active-region magnetic field (the small scale flux required for the surface-cooling 
is a consequence of the decay of active-regions). Observations show that 
the low-latitude branch of torsional-oscillations starts around one to two years
before the sunspots of the new cycle appear. It is possible that magnetic flux
rises towards the surface without forming sunspots in the beginning of a cycle
providing enough small-scale magnetic field, this is however currently neither
confirmed nor ruled out by observations. Alternative explanations for the low-latitude branch of torsional oscillations such as the models of \inlinecite{Schussler1981}, \inlinecite{Yoshimura1981}, and \inlinecite{ Covas2000} are based on the longitudinal component of the Lorentz force. Recently \cite{Rempel2007} showed that torsional-oscillations forced that way are close to the Taylor-Proudman state (alignment of phase with the axis of rotation), which contradicts observations. In addition, the resulting meridional surface-flow has the wrong sign (active region belt outflow). Despite some shortcomings, the model of \inlinecite{Spruit2003} is currently the only proposed explanation that is consistent with the observed meridional and zonal-flow variations close to the solar surface.

\begin{figure}[]
  \centering
 \includegraphics[width=11.cm]{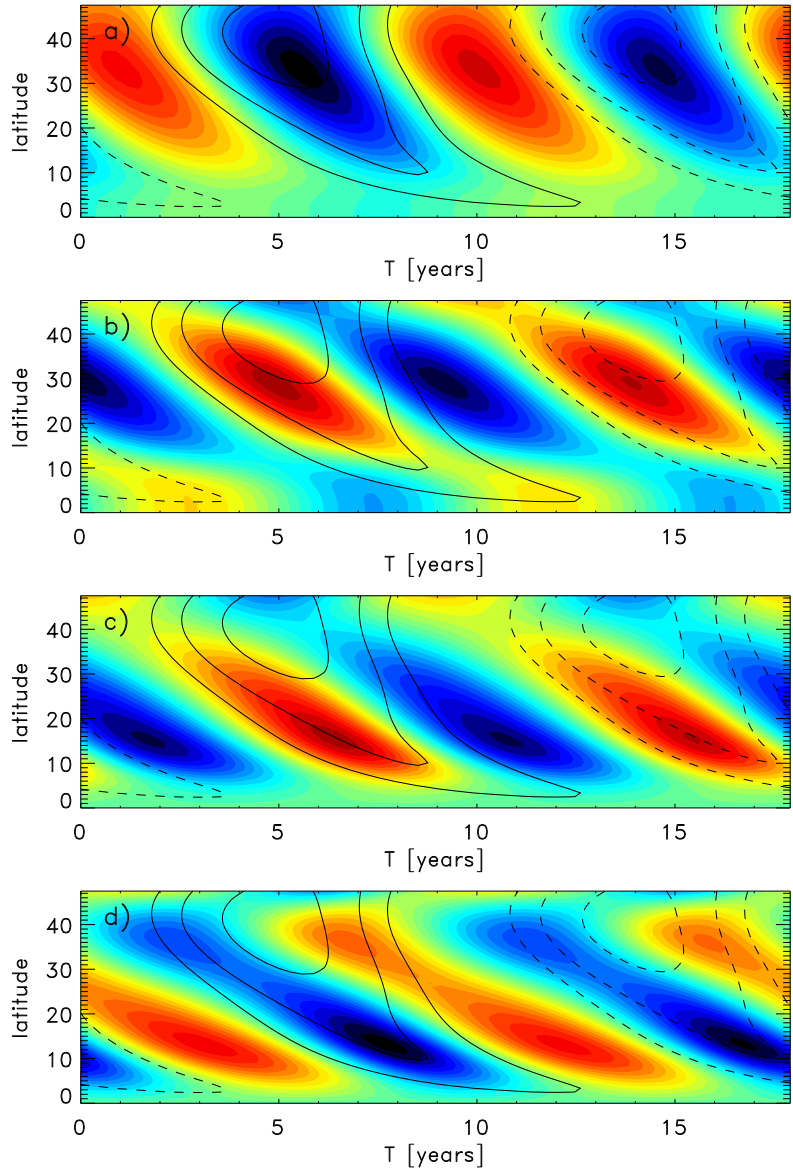} 
\vspace{0.5cm}
\caption{Model results. (a) Surface temperature variation ({\it blue}: cold, {\it red}: hot, amplitude: $0.2$ K). (b) Torsional oscillations ({\it blue}: slower, {\it red}: faster rotation, amplitude: $1.35$ nHz). (c) Meridional flow variation at 
$r=0.985\,R_{\odot}$ ({\it blue}: equatorward, {\it red}: poleward motion, amplitude: 
$2.3$~\ms. (d) Meridional flow variation at 
$r=0.93\,R_{\odot}$ ({\it blue}: equatorward, {\it red}: poleward motion, amplitude: $0.22$~\ms). The variation of the meridional flow pattern 
at $r=0.985\,R_{\odot}$ is almost in anti-correlation with the flow at
$r=0.93\,R_{\odot}$ ($\approx 50$~Mm depth). In all four panels the {\it contour lines}
indicate the butterfly diagram computed from the toroidal field at the base of
the convection zone.
 }
\label{fig:model}
\end{figure}

Figure \ref{fig:model} summarizes the results
of the model in latitudes below $45^\circ$. Figure~\ref{fig:model}a shows the
temperature fluctuation (color shades) caused by increased surface cooling in
the active region belt. The contour lines indicate the magnetic butterfly 
diagram computed from the toroidal field at the base of the convection zone in
the model. Figure~\ref{fig:model}b shows cycle variations of the angular 
velocity (torsional oscillations) and Figure~\ref{fig:model}c shows cycle variations
of the horizontal meridional flow velocity. At the equatorward side of the 
active region belt (indicated by the butterfly diagram) the rotation rate is
increased, which is consistent with the increased poleward meridional flow
transporting material toward the axis of rotation. On the poleward side
of the active region belt the rotation rate is lower, while the meridional-flow perturbation is equatorward. At a depth of around $50$ Mm (Figure \ref{fig:model}d) the meridional-flow perturbation is almost anti-correlated
to the surface flow (active-region belt outflow), indicating that the surface 
cooling drives a flow system that closes in the upper third of the convection 
zone. The flow amplitude at a depth of $50$ Mm is around one order of magnitude lower compared to the surface flow due to the significant increase in density.

\section{Discussion}
\label{sec.discussion}
We presented a model that is qualitatively consistent with the 
observations, in particular the phase of the solar-cycle variations
of the flows. Since the surface-cooling 
term is parametrized, the model can only make predictions about the 
relative amplitude of zonal and meridional flows at different depths, and not 
about the absolute values.  Near the surface, the model is in 
agreement with the data: the relative amplitudes of the torsional oscillation and 
the time-varying component of the meridional flow and their relative phase are 
reproduced well. Deeper in the interior, it appears that 
the model underestimates the amplitude of the time variations of the meridional
flow by an order of magnitude; however, the flow variation is in antiphase to
the surface flow as seen in the data. The lower velocity at depth in the 
dynamo model is a consequence of mass conservation (strong increase in 
density with depth). The much larger outflow which is observed in the data 
cannot be balanced by an inflow close to the surface unless the outflow is 
confined to a very narrow layer.

Overall, it is fair to say that the 
model is encouraging. Alternative models explaining low-latitude torsional
oscillations through the action of the longitudinal Lorentz force tend to
produce zonal flow patterns in the Taylor-Proudman state \cite{Rempel2007}
and meridional outflows rather than inflows close to the surface. On the
other hand, zonal flows appear during solar minimum when no
active regions are present. This requires, in the case of thermal forcing,
that a sufficient amount of magnetic flux is present in the form of small scale
flux elements not evident in synoptic magnetograms 
(see {\it e.g.}, Section 6.3 \opencite{Spruit2003}).  
Also a local treatment of the regions of strong magnetic-field concentrations 
(sunspots and active regions) might be necessary to obtain a better match 
between the model and the data. 

On the observational side, it 
would be useful to invert the travel-time measurements in order to obtain improved  and more reliable estimates of the depth variations of the flows.

\begin{acks}
The National Center for Atmospheric Research is sponsored by the National Science Foundation. M. Rempel thanks Prof. Sch\"{u}ssler and the Max-Planck-Institut f\"{u}r Sonnensystemforschung for their hospitality.
\end{acks}

\end{article} 
\end{document}